\documentclass{appolb}
\usepackage{graphicx}
\usepackage{amsmath}
\usepackage{bm} 

\allowdisplaybreaks[1]
\def\be#1\ee{\begin{equation}#1\end{equation}}
\def\bal#1\eal{\begin{align}#1\end{align}}
\def\bat#1\eat{\begin{alignat}{2}#1\end{alignat}}
\def\bmu#1\emu{\begin{multline}#1\end{multline}}
\def\bga#1\ega{\begin{gather}#1\end{gather}}
\newcommand{\ba}{\begin{array}}
\newcommand{\ea}{\end{array}}
\newcommand{\n}{\notag}

\renewcommand{\d}{\partial}

\begin{document}
\title{The infrared fixed point of Landau gauge \\ Yang-Mills theory%
\thanks{Presented at the conference Light Cone 2012, ``Modern approaches to 
nonperturbative gauge theories and their applications'', in Cracow, Poland.}%
}
\author{Axel Weber
\address{Instituto de F\'{\i}sica y Matem\'aticas \\
Universidad Michoacana de San Nicol\'as de Hidalgo \\
Edificio C-3, Ciudad Universitaria, A. Postal 2-82 \\
58040 Morelia, Michoac\'an, Mexico}
}
\maketitle
\begin{abstract}
Over the last decade, the infrared behavior of Yang-Mills theory in the Landau 
gauge has been scrutinized with the help of Dyson-Schwinger equations and 
lattice calculations. In this contribution, we describe a technically simple 
approach to the deep infrared regime via Callan-Symanzik renormalization group 
equations in an epsilon expansion. This approach recovers, in an analytical 
and systematically improvable way, all the solutions previously found as 
solutions of the Dyson-Schwinger equations and singles out the solution 
favored by lattice calculations as the infrared-stable fixed point (for 
space-time dimensions above two).
\end{abstract}

\section*{}

After almost 40 years of Quantum Chromodynamics, it appears there finally is
significant progress in achieving an analytical or 
semi-ana\-lyt\-ic\-al description of the deep infrared (IR) regime, at least 
in the pure Yang-Mills (YM) sector, i.e., in the absence of dynamical quarks. 
Maybe surprisingly, the methods employed, mainly Dyson-Schwinger (DS) equations
 \cite{SHA97,SHA98,FA02,Zwa02,LS02,FMP09} (and functional renormalization 
group equations \cite{FMP09,PLN04}), are standard methods in quantum field 
theory to go beyond perturbation theory. More recently, it has even been
shown that perturbation theory is successful in describing the IR
behavior of the propagators in the Landau gauge if a mass term for the 
gluons is taken into account \cite{TW10,TW11}.

In this contribution, we will employ renormalization group (RG) methods in 
order to reproduce the solutions of the DS equations and obtain
deeper insight into the IR regime of YM theory. Among other
things, we will find a natural explanation for the success of perturbation 
theory. The main results presented here have been derived in Ref.\
\cite{Web12a}.

We start with the standard formulation of YM theory in the Landau
gauge, with the Nakanishi-Lautrup field to implement the gauge fixing
condition and ghost fields for the local expression of the Faddeev-Popov
determinant leading to the action
\be
S_{\text{FP}} = \int d^D x \left( \frac{1}{4} F^a_{\mu\nu} F^a_{\mu\nu} 
+ i B^a \d_\mu A^a_\mu + \d_\mu \bar{c}^a D^{ab}_\mu c^b \right) 
\label{fadpop}
\ee
in $D$-dimensional Euclidean space-time. The existence of gauge copies in
the Landau gauge \cite{Gri78}, i.e., 
the existence of different gauge equivalent fields that 
all satisfy the gauge condition $\d_\mu A^a_\mu = 0$, forces one to restrict 
the functional integral over the gauge field to the first Gribov 
region.\footnote{Properly, the functional integral should be restricted
to the fundamental modular region. Zwanziger has argued that this further
restriction should have no effect on the correlation functions \cite{Zwa04}.}
It has been observed that this restriction of the functional integral
breaks the BRST invariance of the theory \cite{Zwa93,DGS08}. From an 
RG viewpoint it is natural to expect, and has indeed 
been confirmed in Ref.\ \cite{DGS08}, that quantum corrections then generate 
a gluon mass term of the form
\be
\int d^D x \, \frac{1}{2} A_\mu^a \, m^2 A_\mu^a \,. \label{massterm}
\ee

For the following description of the IR regime of the theory, we
add the latter term to the action \eqref{fadpop}. For small momenta 
$p^2 \ll m^2$, the mass term dominates over the other term in the action 
\eqref{fadpop} that is quadratic in $A_\mu^a$. In our
IR analysis, we will hence effectively replace in the action
\eqref{fadpop} the term quadratic in $A_\mu^a$ by the mass term 
\eqref{massterm}.

We will now perform Wilsonian RG transformations on
this modified action, starting with the simple case of the ``free'' action
$g = 0$ (without interaction terms). The only nontrivial step in the
transformations is the rescaling of the fields. In order to keep the mass 
term \eqref{massterm} invariant, the gauge field has to transform as
\be
A_\mu^a (x) \to s^{D/2} A_\mu^a (s x)
\ee
under a rescaling $x \to x/s$ of the space-time coordinates ($s > 1$). The
scaling dimension of the gluon field is hence $D/2$ instead of the 
canonical value $(D/2) - 1$. The reason is that the scaling dimension of
the gluon field in our case refers to the high-temperature fixed point 
rather than the usual critical fixed point (see, e.g., Ref.\ \cite{Bel91}). 
The scaling dimension of the ghost field has the canonical value.

We now turn to the interacting theory in the vicinity of the fixed
point of the free theory, i.e., for small $g$. The scaling dimensions of
gluon and ghost fields lead to the scaling
\be
g_{\bar{c}Ac} \to s^{1 - (D/2)} g_{\bar{c}Ac}
\ee
of the ghost-gluon coupling constant. Consequently, the ghost-gluon coupling
is relevant for dimensions $D < 2$ and irrelevant for $D > 2$. The three-
and four-gluon couplings are always irrelevant (for $D > 0$). The upper
critical dimension of the theory is then $2$. We will implement an
epsilon expansion around $D = 2$ and neglect the three- and four-gluon
couplings. The IR limit of the theory is then described by the action
\be
S_{\text{GD}} = \int d^D x \left( \frac{1}{2} A_\mu^a \, m^2 A_\mu^a
+ i B^a \d_\mu A^a_\mu + \d_\mu \bar{c}^a D^{ab}_\mu c^b \right) 
\label{ghostdom}
\ee
in $D = 2 + \epsilon$ dimensions.
Note that neglecting the three- and four-gluon couplings is equivalent to
ghost dominance in the sense that to a given perturbative order only the
diagrams with the biggest number of internal ghost propagators are retained.
We will therefore refer to Eq.\ \eqref{ghostdom} as the ghost dominance 
approximation to the theory.

Standard renormalization of the theory defined by Eq.\ \eqref{ghostdom}
leads to the beta function for the dependence of the dimensionless (with 
respect to the correct scaling dimensions of the fields) renormalized 
ghost-gluon coupling constant $\bar{g}_R$ on the renormalization scale $\mu$. 
To order $\epsilon$, one obtains
\be
\beta (\epsilon, \bar{g}_R) = \mu^2 \frac{d}{d \mu^2} \, \bar{g}_R
= \frac{1}{2} \bar{g}_R \left( \frac{\epsilon}{2} - \frac{1}{2}
\frac{N \bar{g}_R^2}{4 \pi} \right) \label{beta}
\ee
where $N$ is the number of colors, see Ref.\ \cite{Web12a} for all details.
Note that we are using the epsilon expansion \emph{above} the critical 
dimension where the theory is, in the usual sense, perturbatively 
nonrenormalizable.

The beta function \eqref{beta} has two fixed points: a trivial IR
attractive one and a nontrivial IR unstable one at $N \bar{g}_R^2/
4 \pi = \epsilon$ (for $\epsilon > 0$). For the IR unstable (and thus
physically irrelevant) one, the solution of the Callan-Symanzik equations
for the two-point functions yields \emph{exactly} one of the two scaling
solutions of the DS equations \cite{Zwa02,LS02}, the one 
usually considered not physical. For the IR attractive (and hence
physical) fixed point, we obtain the decoupling solution of the 
DS equations \cite{AN04,BBL06,Fra08}. The fact that this
fixed point is trivial implies the applicability of perturbation theory
(with the inclusion of a gluon mass term) which has been demonstrated in 
Refs.\ \cite{TW10,TW11}.

We can integrate Eq.\ \eqref{beta} for the coupling constant to obtain 
information on the approach to this trivial fixed point. Solving the
Callan-Symanzik equations for the renormalized propagators with the solution 
$\bar{g}_R (\mu)$ obtained from Eq.\ \eqref{beta} yields
\bal
\big\langle A_{R,\rho}^a (p) A_{R,\sigma}^b (-q) \big\rangle
&= \frac{1}{m^2} \, \frac{1 + (p^2/\Lambda^2)^{\epsilon/2}}
{1 + (\mu^2/\Lambda^2)^{\epsilon/2}} 
\left( \delta_{\rho \sigma} - \frac{p_\rho p_\sigma}{p^2} 
\right) \delta^{ab} (2 \pi)^D \delta (p - q) \,, \n \\
\big\langle c_R^a (p) \bar{c}_R^b (-q) \big\rangle 
&= \frac{1}{p^2} \, \frac{1 + (\mu^2/\Lambda^2)^{\epsilon/2}}
{1 + (p^2/\Lambda^2)^{\epsilon/2}} \, 
\delta^{ab} (2 \pi)^D \delta (p - q) \,. \label{decoupl}
\eal
Here, $\Lambda$ is the characteristic scale where 
$N \bar{g}_R^2 (\Lambda)/4 \pi = \epsilon/2$.

We can compare the IR behavior \eqref{decoupl} directly with
lattice calculations in Landau gauge in the limit of the lattice
parameter $\beta \to 0$. In this limit, the gluonic part in Eq.\
\eqref{fadpop} is completely absent from the action which precisely
corresponds to the ghost dominance approximation \eqref{ghostdom}
(remember that the gluon mass term originates from the BRST symmetry
breaking due to the restriction to the Gribov region). Lattice calculations
at $\beta = 0$ in $D = 3$ and $4$ dimensions ($\epsilon = 1$ and $2$)
nicely confirm the qualitative behavior \eqref{decoupl} \cite{CM10}.

For the full theory ($\beta > 0$), the linear rise with $|p|$ of the
gluon propagator in three dimensions as predicted in Eq.\ \eqref{decoupl}
is also clearly seen in lattice simulations \cite{CM09}. In four dimensions,
however, it turns out that the $p$-dependence of the gluon propagator
generated by the RG improvement in Eq.\ \eqref{decoupl} 
is of the same order in $p^2/m^2$ as the contribution of the term quadratic
in the gluon field in Eq.\ \eqref{fadpop} that we have neglected in our
IR analysis. We therefore have to reestablish the latter term
(it does not receive quantum corrections to one-loop order) with the
result that
\bmu
\big\langle A_{R,\rho}^a (p) A_{R,\sigma}^b (-q) \big\rangle \\
= \left( p^2 + \frac{m^2 (\mu^2 + \Lambda^2)}{p^2 + \Lambda^2} \right)^{-1}
\left( \delta_{\rho \sigma} - \frac{p_\rho p_\sigma}{p^2} 
\right) \delta^{ab} (2 \pi)^D \delta (p - q)
\emu
for $\epsilon = 2$.
The same form of the gluon propagator has been found in Ref.\ \cite{DGS08}
(when the $\langle A_\mu^a A_\mu^a \rangle$-condensate is not taken into
account) and is also qualitatively confirmed by lattice calculations
in $D = 4$ dimensions \cite{BIM07,CM07,SSL07}. Reintroducing the same term
into the action in $D = 3$ dimensions also improves the concordance with the 
lattice calculations \cite{CM09} for the gluon propagator.

From the beta function \eqref{beta} we read off that the nontrivial 
(IR repulsive) fixed point is ultraviolet (UV) attractive. The fact that the
corresponding scaling behavior arises in the UV regime of the ghost 
dominance approximation to the theory, or at lattice parameter 
$\beta = 0$, has lead to some confusion in the literature 
\cite{CM10,MPS10,SS10}. The limit $\beta \to 0$ has been associated with 
the IR regime of YM theory, and hence the appearance of
scaling behavior was interpreted as evidence for the existence of
the scaling solution in IR YM theory (see also Ref.\
\cite{Alk11}). Unfortunately, the values of the exponents of this scaling 
solution have not been firmly established. Our prediction for these values
can be read off from the UV limit $p^2 \gg \Lambda^2$ in Eq.\ 
\eqref{decoupl} (to order $\epsilon$).
It should be clear that the scaling behavior is exclusively related to
the UV stable fixed point of the ghost dominance approximation and
cannot appear in the full YM theory since the latter has a
different UV stable fixed point, the well-known trivial fixed
point associated with asymptotic freedom (while the ghost dominance 
approximation is only valid in the IR regime of YM theory).

Finally, we remark that for $D = 2$ dimensions ($\epsilon = 0$) the two
fixed points in Eq.\ \eqref{beta} coincide and the resulting trivial
fixed point is IR unstable. Consistently, lattice calculations in
two dimensions \cite{Maa07} do not find decoupling behavior for the propagators
(but rather scaling behavior; however, we have not yet been able to
construct the corresponding IR stable fixed point).

For reasons of space, we can only briefly comment that the other
scaling solution of the DS equations [cf.\ the discussion
after Eq.\ \eqref{beta}] also arises in our RG approach if one
implements the so-called horizon condition \cite{Zwa93} in its simplest 
form as IR divergence of the ghost dressing function. The RG analysis can
then be carried through as before. However, this second scaling solution 
turns out to be IR unstable, too, unless the horizon condition is enforced. 
For further details, the reader is refered to Ref.\ \cite{Web12b}. 

In summary, a renormalization group analysis gives considerable insight
into the deep infrared regime of Yang-Mills theory when a gluonic mass
term from the breaking of BRST symmetry due to the restriction of the
functional integral to the first Gribov region is taken into account.
All the different solutions of the Dyson-Schwinger equations are reproduced,
but only the decoupling solution is found to be infrared stable, in
agreement with the results of lattice simulations. The application of
perturbation theory is justified through the triviality of the stable 
fixed point.

%


\subsection*{Acknowledgments}
The author would like to thank the organizers of the conference for an
enjoyable meeting in the beautiful surroundings of Cracow. The present
work was supported by CIC-UMSNH and Conacyt project CB-2009/131787.

\end{document}